\documentclass[hyper,11pt,paper]{article}
\usepackage{jheppub}
\usepackage[usenames,dvipsnames,table]{xcolor}
\usepackage{graphicx,amsmath,amssymb,multirow,array,bm,mathrsfs,bbold,bbm}
\usepackage{epsf,amsfonts,slashed}
\usepackage[numbers,sort&compress]{natbib}
\usepackage[bbgreekl]{mathbbol}
\usepackage{soul,bbold}
\newcommand{\be}{\begin{equation}}
\newcommand{\ee}{\end{equation}}
\newcommand{\ba}{\begin{eqnarray}}
\newcommand{\ea}{\end{eqnarray}}
\newcommand{\beq}{\begin{equation}}
\newcommand{\eeq}{\end{equation}}

\newcommand{\obar}[1]{\overline{#1}}
\newcommand{\wt}[1]{\widetilde{#1}}

\newcommand{\SB}{\text{\bbfamily B}}

\newcommand{\SA}{\text{\bbfamily A}}

\newcommand{\btheta}{\bar{\theta}}
\newcommand{\super}[1]{\mathbb{#1}}

\newcommand{\ssigma}{\bbsigma}

\title{
An entropy current in superspace
}

\author[a]{Kristan Jensen,}
\author[b]{Raja Marjieh,}
\author[c]{Natalia Pinzani-Fokeeva,}
\author[b]{and Amos Yarom}
\affiliation[a]{Department of Physics and Astronomy, San Francisco State University, San Francisco, CA 94132, USA}
\affiliation[b]{Department of Physics, Technion, Haifa 32000, Israel}
\affiliation[c]{Institute for Theoretical Physics, KU Leuven Celestijnenlaan 200D, Leuven B-3001, Belgium}

\emailAdd{kristanj@sfsu.edu}
\emailAdd{sraja@campus.technion.ac.il}
\emailAdd{natascia.pinzanifokeeva@kuleuven.be}
\emailAdd{ayarom@physics.technion.ac.il}

\abstract{ We construct an entropy current using a supersymmetric formulation of the low-energy effective action for the Schwinger-Keldysh generating functional. We define an entropy current quantum mechanically by coupling it to an external source. It is given by the bottom component of an entropy current superfield which is conserved in superspace, but when restricted to real space satisfies a non-conservation law. Our analysis is valid in the probe limit which allows us to fully treat quantum fluctuations.
}

\preprint{\today}

\begin{document}

\maketitle

%------------------------------------------------------
\section{Introduction}
%------------------------------------------------------

Entropy is one of the most fundamental concepts in physics. While it is well-defined and intuitive, its effect on physical processes is somewhat and surprising and far reaching. The second law of thermodynamics has repercussions on a broad spectrum of physical phenomena including phase transitions, black holes and  information theory. 
In the context of hydrodynamics, a local version of the second law tightly constrains the transport properties of fluids. 

Relativistic hydrodynamics can be thought of as a low-energy effective description of a many-body system. In the absence of stochastic noise the degrees of freedom of the hydrodynamic theory can be parameterized by a local temperature, $T$, a local velocity field $u^{\mu}$ and local chemical potential $\mu$. The conserved currents of the theory are local functions of the hydrodynamic variables as long as the latter vary slowly in space and time. For instance, in the absence of conserved charges, the energy-momentum tensor satisfies
$
	T^{\mu\nu} = \epsilon(T) u^{\mu}u^{\nu} + P(T) \left(\eta^{\mu\nu} + u^{\mu} u^{\nu} \right) + \mathcal{O}(\partial)
$
where $\mathcal{O}(\partial)$ are corrections involving derivatives of the hydrodynamic variables which are presumably suppressed by powers of the mean free path $\ell_{mfp}$. The parameters $\epsilon$ and $P$ are the thermodynamic energy density and pressure, with $\epsilon$ determined by $\epsilon = T s- P$, with $s$ the entropy density given by $s = \partial P/ \partial T$.

One way to obtain the aforementioned Gibbs-Duhem relation between energy density, pressure and entropy density is to posit the existence of an entropy current $S^{\mu}$  with non-negative divergence
\begin{equation}
\label{E:localsecond}
	D_{\mu} S^{\mu} \geq 0\,,
\end{equation}
such that
\begin{equation}
\label{E:localconstitutive}
	S^{\mu} = s u^{\mu} + \mathcal{O}(\partial)\,.
\end{equation}
These two defining features of the entropy current are sufficient to obtain the Gibbs-Duhem relation, and many other properties of the fluid: positivity of the conductivity and shear viscosity \cite{LL6}, absence of response to thermal gradients \cite{Jensen:2011xb,Banerjee:2012iz,Jensen:2012jh}, and the interrelation between anomalies and hydrodynamics \cite{Son:2009tf} come to mind.

While the role of a local version of the second law, as given in \eqref{E:localsecond}, is intuitively clear, its appearance in any effective theory is unanticipated.  Noether's theorem guarantees that for each symmetry we will have a conserved current whose divergence equals zero and a corresponding Ward identity, but it is difficult to conceive of a mechanism which will lead to an inequality rather than an equality. 

In recent years several proposals were made to identify a symmetry which generates a conserved entropy current in the absence of dissipation, see e.g.~\cite{Haehl:2014zda,Haehl:2015pja,deBoer:2015ija,Crossley:2015tka,Sasa:2015zga}. More recently, the authors of  \cite{Glorioso:2016gsa,Glorioso:2017fpd} have maintained that a positive divergence entropy current can be constructed by appealing to a positivity constraint on the imaginary part of the Schwinger-Keldysh effective action. In more detail, the formalism developed in  
\cite{Haehl:2015foa,Crossley:2015evo,Haehl:2015uoc,Haehl:2016pec,Haehl:2016uah,Glorioso:2016gsa,Jensen:2017kzi,Glorioso:2017fpd,Haehl:2017zac,Geracie:2017uku}
allows one to construct for a low-energy Wilsonian effective action $S_{eff}$ for the Schwinger-Keldysh effective theory. 
Integrating the exponentiated effective action leads to the low-energy Schwinger-Keldysh partition function
\begin{equation}
	Z = \int D\xi \,e^{i S_{eff}} \,,
\end{equation}
with $\xi$ the low energy dynamical degrees of freedom. One difference between ordinary and Schwinger-Keldysh effective field theory is that, here, unitarity does not require $S_{eff}$ to be real. Convergence of the functional integral constrains the imaginary part of $S_{eff}$ to be bounded below. The authors of \cite{Glorioso:2016gsa,Glorioso:2017fpd} further showed that unitarity implies that $\text{Im}(S_{eff})\geq 0$ and then used this constraint to construct an entropy current with properties \eqref{E:localsecond} and \eqref{E:localconstitutive}. 

In this work we take a somewhat different path and obtain an entropy current by coupling it to an external source.  This procedure allows one to obtain a ``consistent''  super entropy current $\super{S}^{\prime\,I}$ by varying $Z$ with respect to a source $\super{A}_I$. Here $I$ runs over spacetime indices $\mu$ and two superspace indices $\theta$ and $\btheta$ which may be thought of as a useful bookkeeping device which captures the special symmetries associated with the effective action \cite{Haehl:2015foa,Crossley:2015evo,Haehl:2015uoc,Haehl:2016pec,Haehl:2016uah,Jensen:2017kzi,Glorioso:2017fpd,Haehl:2017zac}. The boldface font for $\super{S}^{\prime\,I}$ and $\super{A}_I$ emphasizes that these are functions of both the spacetime coordinates and the superspace coordinates. The consistent entropy current will be conserved in superspace
\begin{equation}
	D_I \super{S}^I=0
\end{equation}
but the spacetime components of its bottom component, $\super{S}^{\mu} =S^{\prime\mu}(x)+\mathcal{O}(\theta,\btheta) $ will satisfy the on-shell relation
\begin{equation}
\label{E:Sconservation}
	D_\mu {S'}^\mu = - S^{\theta}_{\bar{g}} - S^{\btheta}_g \,,
\end{equation}
where $S_{\bar{g}}^{\theta}$ and $S_g^{\btheta}$ are associated with the $\theta$ and $\btheta$ components of the super entropy current. We will refer to $S^{\prime\mu}$ as the consistent entropy current.

We show that in a saddle point approximation, ${S'}^{\mu} = s u^{\mu} + \mathcal{O}(\partial)$, and that the right-hand side of \eqref{E:Sconservation} is constrained to be positive  semidefinite up to a total derivative, which may be made to appear at least at 4th order in the derivative expansion. This feature of ${S'}^{\mu}$ allows one to extract the ``hydrodynamic''  entropy current, $S^{\mu}=S'^{\mu} + O(\partial)$ 
whose divergence is positive semidefinite.

The external field $\super{A}_I$ which naturally couples to $\super{S}'^I$ bears remarkable similarities to a recently proposed (dynamical) gauge field associated with the thermodynamic free energy current \cite{Haehl:2014zda,Haehl:2015pja,Haehl:2015foa,Haehl:2015uoc} (see also the precursor \cite{Jensen:2013rga}). One distinction between our construction and previous ones is that the superfield $\super{A}_I$ is an external source in our setup and not a dynamical variable. Thus, $\super{S}^{\prime I}$ may be derived by varying the Schwinger-Keldysh partition function $Z[\super{A}]$ and one may treat $S^{\prime\mu}$ as a definition of a quantum entropy current. Indeed, one may, for instance, use 
our effective action to compute correlation functions of the entropy current.

The remainder of this work is organized as follows. In Section \ref{S:effectiveaction} we remind the reader of the formalism of \cite{Jensen:2017kzi} used to construct $S_{eff}$. We focus on a probe limit which only includes the dynamics of $U(1)$ currents at low chemical potential. In Section \ref{S:entropy} we construct an entropy current in superspace and show that the right-hand side of \eqref{E:Sconservation} is positive definite in a gradient expansion up to a total derivative, leading to an exact definition for $S^{\mu}$. In Section \ref{S:newentropy} we demonstrate the second law for the total entropy independently of a derivative expansion.

%------------------------------------------------------
\section{The hydrodynamic effective action}
\label{S:effectiveaction}
%------------------------------------------------------

Let us begin by recalling the basic ingredients needed to construct the Schwinger-Keldysh hydrodynamic effective action $S_{eff}$.\footnote{This action accounts for the low-energy physics of the Schwinger-Keldysh partition function
\beq
Z[A_1,A_2] = \text{Tr}\left( \mathcal{U}[A_1] e^{-\beta H}\mathcal{U}_2^{\dagger}[A_2]\right)\,,
\eeq
where $\mathcal{U}[A]$ is the time evolution operator from the infinite past to the infinite future in the presence of a source $A$ and $e^{-\beta H}$ is the thermal density matrix of the initial state.} As is the case for any effective theory, $S_{eff}$ is the most general action one can construct which is compatible with the symmetries  of the problem. The relevant symmetries involve 
\begin{enumerate}
\item
	
	A doubling of the symmetries associated with the doubled external sources.
\item
	A reality condition on $S_{eff}$.
\item
	A topological symmetry associated with the vanishing correlation functions of ``difference operators.'' (We refer to this as the Schwinger-Keldysh symmetry.)
\item 
	A non-local $\mathbb{Z}_2$ symmetry which we refer to as the full KMS symmetry. 
\end{enumerate}
The dynamical degrees of freedom of the theory may be thought of as doubled embedding functions of a Lagrangian description of the fluid. The reader is referred to \cite{Haehl:2015foa,Crossley:2015evo,Haehl:2015uoc,Haehl:2016pec,Haehl:2016uah,Jensen:2017kzi,Glorioso:2017fpd,Haehl:2017zac} for an extensive discussion. 

In this work we will closely follow \cite{Jensen:2017kzi}. There, following previous work, three of us argued for the following construction for a Schwinger-Keldysh effective action for hydrodynamics. The topological symmetry is enforced by adding ghost degrees of freedom and then imposing two BRST-like symmetries $Q$ and $\overline{Q}$ (which are exchanged by the full KMS symmetry). The degrees of freedom are embedding functions ${X}_1^{\mu}$, $X_2^{\mu}$ which serve as dynamical mappings from a ``worldvolume'' with coordinates $\sigma^i$ to two ``target spaces,'' and their associated ghosts $X_g^{\mu}$ and $X_{\bar{g}}^{\mu}$. For charged matter one has, in addition, phases $C_1$, $C_2$ and ghost fields $C_{g}$ and $C_{\bar{g}}$. Apart from the dynamical fields the action depends on external metrics $g_{1\,\mu\nu}$ and $g_{2\,\mu\nu}$ , as well as external flavor fields $B_{1\,\mu\nu}$ and $B_{2\,\mu\nu}$. The action also depends on the thermodynamic parameters of the initial state. These are characterized by a timelike vector $\beta^i$ and a flavor gauge transformation parameter $\Lambda_{\beta}$.\footnote{\label{F:staticgauge} Note that we can always pick a ``static gauge'' for the worldvolume coordinates and flavor gauge such that $\beta^i \partial_i = \beta \partial_0$ and $\Lambda_{\beta}=0$, with $\beta$ the inverse temperature of the thermal state in the infinite past. In this gauge our expressions are closely related to those in \cite{Crossley:2015evo,Glorioso:2017fpd}.}

Further following \cite{Jensen:2017kzi},  we will restrict ourselves to a probe limit wherein we are considering the dynamics of a sufficiently weak conserved charge propagating in a fixed thermally equilibrated background. Working in a flat target space, this implies that $g_{1\,\mu\nu}= g_{2\,\mu\nu}=\eta_{\mu\nu}$ and that the $X$'s are no longer dynamical and take on their value at equilibrium, 
\begin{equation}
\label{E:probeX}
	X_1^{\mu} = X_2^{\mu} = X^{\mu}_{eq} \equiv \sigma^i \delta_i^{\mu}. 
\end{equation}	
The only remaining dynamical degrees of freedom in our setup are the $C$'s and their ghost partners whose dynamical equations relate to charge conservation
\begin{align}
\begin{split}
	D_{\mu}J^{\mu}_1 =0\,,
	\qquad
	D_{\mu}J^{\mu}_2 =0\,,
\end{split}
\end{align}
and we consider configurations where the currents are perturbatively small. In this limit the two stress tensors  approximately coincide,
\beq
T_1^{\mu\nu} = T_2^{\mu\nu} = T_{eq}^{\mu\nu}\,, 
\eeq
and are conserved,
\begin{equation}
\label{E:stressconservation}
	D_{\mu}T_{eq}^{\mu\nu}=0\,.
\end{equation}	
At this point we emphasize that even though there is a nonzero current, by assumption it is sufficiently weak so that the Joule heating term in the Ward identity can be neglected. That is, \eqref{E:stressconservation} is satisfied instead of $D_{\mu}T^{\mu\nu} = G^{\nu\mu}J_{\mu}$ with $G$ the external field strength and $J_{\mu} $ the physical $U(1)$ current.

As we already mentioned, there are nilpotent supercharges $Q$ and $\overline{Q}$ satisfying $Q^2=\overline{Q}^2=0$. They satisfy the algebra
\begin{equation}
\label{E:commrelation}
	\{Q,\,\overline{Q} \} = i \delta_{\beta}
\end{equation}
where $\delta_{\beta}$ acts as a combination of a Lie derivative in the $\beta$ direction and, in addition, as a flavor gauge transformation with parameter $\Lambda_{\beta}$ when acting on connections, or objects which are charged under the flavor symmetry. For example,
\begin{equation}
\label{E:deltabeta}
	\delta_{\beta}\phi = \beta^i \partial_i \phi
\end{equation}
where $\phi$ is a neutral scalar.\footnote{In the static gauge we would have, $\delta_{\beta}\phi = \beta \partial_{0}\phi$.}
In order to implement these symmetries we add fictitious superspace coordinates to the spacetime which we denote by $\theta$ and $\btheta$.

The dynamical fields can be collected into superfields
\begin{equation}
\label{E:longmultiplet}
	\super{C} = \mathcal{R} C_r + \theta  C_{\bar{g}} + \btheta C_g + \btheta \theta  \mathcal{A} C_a\,,
\end{equation}
where $C_r = (C_1+C_2)/2$, $C_a=C_1-C_2$, and $\mathcal{A}/ \mathcal{R} = \coth(i \delta_{\beta}/2) i \delta_{\beta}/2 $.\footnote{
More generally, there is some freedom in the ghost terms of the dynamical fields which one can parameterize in the following way:
\begin{equation}
	\super{C} = \mathcal{R} C_r + \theta \overline{\mathcal{G}} C_{\bar{g}} + \btheta \mathcal{G} C_g + \btheta \theta  \mathcal{A} C_a\,,
\end{equation}
and there is a somewhat involved expression for $\mathcal{G}$ and $\overline{\mathcal{G}}$. As discussed in \cite{Jensen:2017kzi} we may set $\mathcal{G}= \overline{\mathcal{G}} = 1$ which is what we have done in \eqref{E:longmultiplet}. The exact values of $\mathcal{G}$ and $\overline{\mathcal{G}}$ will not play a role in what follows.} 
For every multiplet of the type \eqref{E:longmultiplet} there exists a tilde'd multiplet
\begin{equation}
\label{E:tildedmultiplet}
	\wt{\super{C}} = \mathcal{R} \wt{C}_r + \theta \wt{C}_{\bar{g}} + \btheta \wt{C}_g + \btheta \theta   \mathcal{A} \wt{C}_a\,,
\end{equation}
with 
\begin{align}\label{E:def}
\begin{split}
	\wt{C}_r &=  \frac{1}{2} \left( 1+ e^{-i\delta_{\beta}} \right) C_r + \frac{1}{4}  \left( 1- e^{-i\delta_{\beta}} \right) C_a\,,
	\qquad
	\wt{C}_g = \frac{2}{1+e^{i\delta_{\beta}}} C_g\,,  \\
	\wt{C}_{a} &=  \frac{1}{2} \left(1 + e^{-i\delta_{\beta}} \right) C_a + \left(1-e^{-i\delta_{\beta}} \right) C_r\,,
	\hspace{.44in}
	\wt{C}_{\bar{g}} =  \frac{e^{-i\delta_{\beta}} + 1}{2}  C_{\bar{g}}\,.
\end{split}
\end{align}

With these definitions the action of the supercharges on the above superfields is given by
\begin{align}
\begin{split}
	\delta_{Q} \super{C} &= \frac{\partial}{\partial \theta} \super{C} \,,
	\hspace{1in}
	\delta_{\overline{Q}} \super{C} = \left(\frac{\partial}{\partial \btheta} + i \delta_{\beta} \theta \right) \super{C} \,, \\
	\delta_{Q} \wt{\super{C}}& = \left(\frac{\partial}{\partial \theta} + i \delta_{\beta} \btheta \right) \wt{\super{C}} \,,
	\qquad
	\delta_{\overline{Q}} \wt{\super{C}} =\frac{\partial}{\partial \btheta} \wt{\super{C} } \,.
\end{split}
\end{align}
The associated superderivatives which anticommute with $Q$ and $\overline{Q}$ are given by 
\begin{align}
\begin{split}
	{D}_{\theta}  \super{C} &= \left(\frac{\partial}{\partial \theta} - i \delta_{\beta} \btheta \right) \super{C} \,,
	\qquad
	{D}_{\btheta} \super{C} = \frac{\partial}{\partial \btheta} \super{C} \,, \\
	\wt{{D}}_{\theta} \wt{\super{C}} &=\frac{\partial}{\partial \theta} \wt{\super{C}}  \,,
	\hspace{1in}
	\wt{{D}}_{\btheta}  \wt{\super{C}} = \left(\frac{\partial}{\partial \btheta} - i \delta_{\beta} \theta \right) \wt{\super{C} }\,.
\end{split}
\end{align}

In order to construct a gauge invariant action we join the dynamical fields with the sources so that the resulting object is invariant under target space gauge transformations. We denote the resulting supermultiplet by 
\begin{equation}	
\label{E:superB}
	\super{B}_i = \mathcal{R} B_{r\,i} + \btheta\theta \mathcal{A} B_{a\,i} + \partial_i \super{C}\,,
\end{equation}
where 
\begin{align}
\begin{split}
\label{eq:embeddings}
	B_{r\,i} &= \frac{1}{2} \left( \partial_i X_1^{\mu}  B_{1\,\mu}(X_1) + \partial_i X_2^{\mu} B_{2\,\mu}(X_2) \right) \\
	B_{a\,i} &=  \left( \partial_i X_1^{\mu} B_{1\,\mu}(X_1) - \partial_i X_2^{\mu} B_{2\,\mu}(X_2) \right)\,,
\end{split}
\end{align}
and the $X_1^{\mu}$ and $X_2^{\mu}$ are given by their equilibrium value \eqref{E:probeX}.
Had we not been working in the probe limit, we would have been compelled to construct a worldvolume supermetric $\super{g}_{ij}$ which contains the embedding functions and the metrics of the target-space. In our probe limit we have $\super{g}_{ij} =  \eta_{\mu\nu} \delta^{\mu}_i \delta^{\nu}_j$.

The most general effective action, $S_{eff}$, constructed out of these fields and which satisfies the required symmetries is 
\begin{multline}
\label{E:fullaction}
	S_{eff} = \frac{1}{2} \int d^d\sigma d\theta d\btheta  \sqrt{-\super{g}}\, L(\super{B}_i,\,\super{g}_{ij},\,\super{D}_i,\,i {D}_{\theta},\,{D}_{\btheta};\,\beta,\,\Lambda_{\beta}) \\
	+\frac{1}{2}  \sqrt{-\wt{\super{g}}}\, L(\eta_B \wt{\super{B}}_i,\,\eta_g \super{g}_{ij},\,\eta_{\partial} \wt{\super{D}}_i,\,-i \wt{{D}}_{\btheta},\,\widetilde{{D}}_{\theta};\,\eta_{\beta} \beta,\,-\Lambda_{\beta})\,.
\end{multline}
Here tilde'd superfields are related to untilde'd ones as in \eqref{E:def} (and we have used $\wt{\super{g}}_{ij} = \super{g}_{ij}$. Also, $\super{D}_i$ denotes the covariant derivative constructed from the metric $\super{g}_{ij}$,   $\eta_B$ and $\eta_g$ are the CPT eigenvalues of $\super{B}$ and $\super{g}$, $\eta_{\partial} \wt{\super{D}}_i$ is a CPT transformation of $\super{D}_i$ and $\eta_{\beta}$ is the CPT eigenvalue of $\beta$. We refer to the second term on the right-hand side of \eqref{E:fullaction} as the KMS partner of the first. 

Often, it is convenient to decompose $L$ such that
\begin{equation}
\label{E:actionexpansion}
	L = L_0(\super{B},\super{D}_i) 
	+  \sum_{n=0} i^{n+1}{L}^{j_1j_2 k_1 \ldots k_n}(\super{B},\super{D}_i)  {D}_{\theta}\super{B}_{j_1} {D}_{\btheta} \super{B}_{j_2} {D} \super{B}_{k_1} \ldots {D} \super{B}_{k_n}  
	+ \hbox{ghost terms}
\end{equation}
where, 
\begin{equation}
	{D} = {D}_{\btheta} {D}_{\theta}\,,
\end{equation}
and by ``ghost terms'' we mean terms which vanish when ghosts are set to zero. We have omitted the explicit dependence on $\beta$, $\Lambda_{\beta}$ and $\super{g}_{ij}$ for brevity. The capitalized Roman indices specify spacetime indices. 
We refer to $L_0$ as the scalar contribution to $L$ and to the $n$-th term in the sum on the right-hand side of \eqref{E:actionexpansion} as the $n+2$-th tensor term in $L$. 

Let us pause to explain what precisely we mean by the probe limit. First, recall that the conserved current of the theory may be computed by varying the Schwinger-Keldysh generating functional $W=-i\ln Z$ with respect to the gauge field $B_{a\,i}=B_{1\,i}-B_{2\,i}$ and then setting $B_{1\,i} = B_{2\,i} \equiv B_{i}$:
\begin{equation}
	\langle J^i\rangle  = \frac{1}{\sqrt{-g}} \frac{\delta W}{\delta B_{a\,i}} \Bigg|_{B_a=0}\,.
\end{equation}
See e.g., \cite{Crossley:2015evo} for a modern presentation of the subject. Had we turned on external metrics we would have been able to similarly obtain the expectation value of the stress tensor. The equations of motion for the $C$'s and $X^{\mu}$'s ensure current conservation $D_{\mu}J^{\mu}=0$ and energy-momentum conservation $D_{\mu}T^{\mu\nu} = G^{\nu\mu}J_{\mu}$ where the right-hand side of the latter equation is referred to as a Joule heating term. 

In the probe limit we introduce a formal expansion parameter $\epsilon$ and take the external fields $B_{1\,\mu}$ and $B_{2\,\nu}$, as well as the superfield $\super{C}$ to be $O(\epsilon)$. In this limit the current is also $O(\epsilon)$. Invariance under charge conjugation, C, is enough to guarantee the validity of the probe limit. It implies that the external fields and $C$'s backreact on the $X$'s at $O(\epsilon^2)$, i.e. the solution to the equations of motion for the $X$'s has the form $X_s^{\mu} = \sigma^i \delta_i^{\mu} + \epsilon^2 \delta X^{\mu}_s + O(\epsilon^4)$ with $s=1,2$. The stress tensors are similarly given by $T^{\mu\nu}_s = T^{\mu\nu}_{eq} + \epsilon^2 \delta T^{\mu\nu}_s + O(\epsilon^4)$. 

In this note we work with the effective action to $O(\epsilon^2)$,
%Thus, we may 
and so neglect terms which are cubic in $\super{B}$ in \eqref{E:actionexpansion} (which implies neglecting terms which are quadratic in the components of $\super{B}$ in the equations of motion). Thus, \eqref{E:actionexpansion} truncates to
\begin{equation}
\label{E:actionexpansionshort}
	L = L_0(\super{B},\super{D}_i) 
	+  i^{}{L}^{ij}(\super{D}_i) {D}_{\theta}\super{B}_i {D}_{\btheta} \super{B}_j  
	+ \hbox{ghost terms}\,.
\end{equation}

In what follows we will assume that there exists a parameter which allows us to take a saddle point approximation. For instance, in large $N$ gauge theories $1/N$ is just such a small parameter. Another saddle point can be obtained in a statistical mechanical limit where quantum fluctuations are suppressed relative to thermal ones. Such an approximation was carried out in \cite{Crossley:2015evo}. We will show how to incorporate such an approximation into our formalism in a future publication \cite{panoply}. After taking the saddle-point approximation one may obtain the constitutive hydrodynamic relations by varying the effective action with respect to the sources $B_{a\,i}$ and pushing forward these relations to the target-space. Since the pushforward will not alter the algebraic structure of these relations we can read them off directly from their worldvolume counterparts. We refer the reader to \cite{Jensen:2017kzi} for details.

%------------------------------------------------------
\section{The entropy current}
\label{S:entropy}
%------------------------------------------------------

 There are a few equivalent versions of the second Law of hydrodynamics used in the literature. The most common one, appearing in e.g. Landau and Lifshitz \cite{LL6}, is that there ought to exist a current $S^{\mu} = s u^{\mu} + O(\partial)$ with $s$ the entropy density which satisfies $D_{\mu}S^{\mu} \geq 0$ for fluid configurations which solve the hydrodynamic equations. In the context of effective field theory for hydrodynamics, this is an ``on-shell'' second Law. There is also an ``off-shell'' second law~\cite{Loganayagam:2011mu}, and elaborated on in \cite{Haehl:2014zda,Haehl:2015pja}, which is the assertion of a current $S^{\mu} = s u^{\mu} + O(\partial)$ which satisfies
\beq
	\label{E:2ndS}
	D_{\mu}S^{\mu} + \beta_{\mu}\left( D_{\nu}T^{\mu\nu} - G^{\mu}{}_{\nu}J^{\nu}\right) + \frac{\mu}{T}D_{\mu}J^{\mu}  =\mathcal{S} \geq 0\,,
\eeq
for any fluid configuration, including those that do not solve the hydrodynamic equations. This off-shell version can be rewritten in terms of a free energy current $N^{\mu}/T = S^{\mu} - T^{\mu\nu}\beta_{\nu} - \frac{\mu}{T}J^{\mu}$, which satisfies
\beq
	\label{E:2ndG}
D_{\mu}\left( \frac{N^{\mu}}{T}\right) - \frac{1}{2}T^{\mu\nu}\delta_{\beta}  g_{\mu\nu} - J^{\mu}\delta_{\beta}B_{\mu} =\mathcal{S}\geq 0\,,
\eeq
for any fluid configuration. 

As a prelude to our construction of the entropy current, let us take a step back and consider an action $S = \int d^d\sigma \sqrt{-g}\, L$ with a dynamical variable $\phi$ and external metric $g_{ij}$. Under a general variation of the quantum and external fields the action varies by
\begin{equation}
\label{E:simplest}
	\delta S = \int  d^d \sigma \sqrt{-g} \left(E \delta \phi + \frac{1}{2}T^{ij} \delta g_{ij} \right)\,,
\end{equation}
with $E$ the equation of motion for $\phi$ and $T^{ij}$ the stress tensor. Now fix a vector field $\beta^i$ and define a transformation $\delta_{\beta}$ which acts as a Lie derivative on $g_{ij}$ and $\phi$.

Let us now consider a ``gauged'' version of the transformation $\delta_{\beta}$, $\delta_T$, which satisfies
\begin{equation}
\label{E:gaugeddeltabeta}
	\delta_{T} \phi = \Lambda_T \delta_{\beta}\phi\,,
	\qquad
	\delta_{T} g_{ij} = \Lambda_T \delta_{\beta} g_{ij}\,,
\end{equation}
with $\Lambda_T$ a spacetime dependent parameter. If we minimally couple the theory to an external flavor field $A_i$,
\begin{equation}
\label{E:DtoDA}
	D_i \to D_{(A)\,i} + A_i \delta_{\beta}\,,
\end{equation}
with $D_{(A)\,i}$ a covariant derivative with connection
\beq
	\label{eq:Crist}
	(\Gamma_{(A)})^i{}_{jk} = \frac{1}{2}g^{il} \left( (\partial_j + A_j \delta_{\beta}) g_{kl} + (\partial_k + A_k \delta_{\beta})g_{jl} - (\partial_l + A_l \delta_{\beta})g_{jk}\right)\,,
\eeq
then requiring that covariant derivatives of fields transform in the same way as fields themselves, e.g.
\beq
	\delta_T \left( D_{(A)\,i} \phi \right)= \Lambda_T \delta_{\beta}\left( D_{(A)\,i}\phi\right)\,,
\eeq
implies that $A_i$ varies under $\delta_T$ as
\begin{equation}
\label{E:Arule}
	\delta_{T} A_i = \Lambda_T \delta_{\beta}A_i - A_i \delta_{\beta} \Lambda_T - \partial_i \Lambda_T\,.
\end{equation}
The Lagrangian density $L$, defined through $S=\int d^d\sigma \sqrt{-g} \, L$, then satisfies
\begin{equation}
	\delta_{T}L = \Lambda_T \delta_{\beta} L\,.
\end{equation}
If we further modify the measure
\begin{equation}\label{eq:measure}
	\sqrt{-g} \to \frac{\sqrt{-g}}{\beta^i A_i+1}\,,
\end{equation}
then the resulting action 
\beq
	S = \int \frac{d^d\sigma \sqrt{-g}}{\beta^i A_i + 1} L\,,
\eeq
is invariant under $\delta_T$
\begin{equation}
	\delta_{T}S = 0\,.
\end{equation}
Thus, if we define the current $S^i$ conjugate to $A_i$ via
\beq
\delta S = \int \frac{d^d\sigma \sqrt{-g}}{\beta^i A_i + 1 }\left( E\delta \phi + \frac{1}{2}T^{ij} \delta g_{ij} - S^i \delta A_i\right)\,,
\eeq
then when $A_i=0$ we obtain the on-shell relation
\begin{equation}
\label{E:Sforphi}
	D_i S^i \big|_{A_i=0} = \frac{1}{2} T^{ij} \delta_{\beta}g_{ij}\,.
\end{equation}
Note that this resembles the Gibbsian version of the second Law~\eqref{E:2ndG}. As one may have expected,
$S^i$ will be conserved on-shell if $\beta$ is a Killing vector, $\delta_{\beta}g_{ij} = 0$. It will coincide with the conserved current associated with the Killing symmetry.

Let us now take a step forward and consider a bosonic sigma model with dynamical degrees of freedom $X^{\mu}$ and $C$, and sources $B_{\mu}$ and $g_{\mu\nu}$ which appear in the action only through their pullbacks
\begin{equation}
	\label{E:pullback}
	B_i=B_{\mu}(X)\partial_{i}X^{\mu}+\partial_{i}C\,,
	\qquad
	g_{ij} = g_{\mu\nu}(X)\partial_i X^{\mu} \partial_j X^{\nu}\,,
\end{equation}
and derivatives thereof. This is not quite the setup we wish to consider, but taking this sidetrack will allow us to motivate our main construction more clearly. We can now define a transformation $\delta_{\beta}$ which generates a worldvolume translation parameterized by a vector $\beta^i$ and gauge transformation parameterized by $\Lambda_{\beta}$,
\begin{equation}
	\delta_{\beta} X^{\mu} = \beta^i \partial_i X^{\mu}\,,
	\qquad
	\delta_{\beta} C = \beta^i \partial_i C + \Lambda_{\beta}\,.
\end{equation}
Note that $g_{\mu\nu}$ and $B_{\mu}$ being functions of the target space are, in this setup, inert under $\delta_{\beta}$.

Using the lessons learnt from gauging the $\delta_{\beta}$ symmetry in \eqref{E:simplest}  we now require
\begin{align}
\begin{split}
	\label{E:U1Tv1}
	\delta_{T}X^{\mu}&=\Lambda_T \delta_{\beta}X^{\mu}\,,
	\\
	 \delta_{T}C&=\Lambda_T \delta_{\beta}C\,,
\end{split}
\end{align}
and take the external fields $g_{\mu\nu}$ and $B_{\mu}$ to be invariant under $\delta_T$. The partial derivatives of $X^{\mu}$ and $C$, which are used to pullback $g_{\mu\nu}$ and $B_{\mu}$, do not transform in the same way as $X^{\mu}$ and $C$. We modify them as
\beq
	\partial_i \to \partial_i + A_i \delta_{\beta}\,,
\eeq
so that the pullbacks of $g_{\mu\nu}$ and $B_{\mu}$ become
\begin{align}\label{E:BtoBA}
\begin{split}
	g^{(A)}_{ij} &= g_{\mu\nu}(X) \left( \partial_i + A_i \delta_{\beta}\right) X^{\mu} \left( \partial_j + A_j \delta_{\beta}\right)X^{\nu} 
	\\
	&= g_{ij} + \beta_i A_j + \beta_j A_i + \beta^2 A_i A_j\,,
	\\
	B^{(A)}_i & = B_{\mu}(X)\left( \partial_i + A_i \delta_{\beta}\right)X^{\mu} + \left( \partial_i + A_i \delta_{\beta} \right)C
	\\
	& = B_i + A_i \nu\,,
\end{split}
\end{align}
where $g_{ij}$ and $B_i$ are the ordinary pullbacks of $g_{\mu\nu}$ and $B_{\mu}$, indices are lowered with $g_{ij}$ and $\nu = \beta^i B_i + \Lambda_{\beta}$. So defined, these pullbacks transform in the same way as $X^{\mu}$ and $C$,
\begin{align}
\begin{split}
\label{E:verifiedsigmatrans}
	\delta_T g_{ij}^{(A)} &= \Lambda_T \delta_{\beta} g_{ij}^{(A)}\,,
	\\
	\delta_T B_i^{(A)} &= \Lambda_T \delta_\beta B_i^{(A)}\,.
\end{split}
\end{align}
The variation of $\sqrt{-g} = \sqrt{-\text{det}(g_{ij})}$ is
\beq
	\delta_T \sqrt{-g} = \frac{1}{2}\sqrt{-g} g^{ij}\delta_T g_{ij} =\Lambda_T \beta^i \partial_i \sqrt{-g} + \sqrt{-g}\delta_{\beta}\Lambda_T
\eeq
and so an action of the form 
\beq
	S = \int d^d\sigma \sqrt{-g} \, L(g^{(A)},B^{(A)})\,,
\eeq
is then invariant, $\delta_T S=0$.

Defining the various currents by
\beq
	\delta S = \int d^d\sigma \sqrt{-g} \left( \frac{1}{2}T^{ij} \delta g_{ij} + J^i \delta B_i - S^i \delta A_i\right)\,,
\eeq
and using~\eqref{E:pullback}, we can rewrite the variation of $S$ as
\beq
	\delta S = \int d^d\sigma \sqrt{-g} \left( \frac{1}{2}T^{\mu\nu} \delta g_{\mu\nu} + J^{\mu}\delta B_{\mu} - \mathcal{E}_{\mu}\delta X^{\mu} - \mathcal{E} \delta C - S^i \delta A_i\right)\,,
\eeq
with $T^{\mu\nu} = T^{ij} \partial_i X^{\mu}\partial_j X^{\nu}$ and $J^{\mu} =J^i\partial_i X^{\mu}$ as well as
\begin{align}
\begin{split}
	\mathcal{E}_{\mu} & = D^{\nu}T_{\mu\nu} - G_{\mu\nu}J^{\nu}\,,
	\\
	\mathcal{E} & = D_{\mu}J^{\mu}\,.
\end{split}
\end{align}
Now using that under $\delta_T$ both $g_{\mu\nu}$ and $B_{\mu}$ are invariant, but $X^{\mu}$ and $C$ vary as~\eqref{E:U1Tv1}, we see that when $A_i = 0$, $\delta_TS = 0$ implies
\beq
	\label{E:bosonicsigma}
	D_i S^i + \beta_{\mu}\left( D_{\nu}T^{\mu\nu} - G^{\mu}{}_{\nu}J^{\nu}\right) + \nu D_{\mu}J^{\mu}= 0\,.
\eeq
This recalls the entropic second Law~\eqref{E:2ndS}.

We could have defined another version of $\delta_T$ under which the sigma model fields $X^{\mu}$ and $C$ were invariant, but $g_{\mu\nu}$ and $B_{\mu}$ varied. The Ward identity of that transformation would resemble the Gibbsian second Law~\eqref{E:2ndG} instead. In what follows, however, it will be more convenient to supersymmetrize the transformation in which the sources are inert and the sigma model fields transform.

Let us now consider the probe limit of the bosonic sigma model described above. In the probe limit we consider a solution where $g_{\mu\nu}=\eta_{\mu\nu}$ and the $X$'s take on their classical value $X^{\mu} = X_{eq}^{\mu} =\delta^{\mu}_{i}\sigma^{i}$. Since the $X^{\mu}$'s take on their classical value we can no longer define a transformation $\delta_T$ under which the $X$'s vary as $\delta_T X^{\mu} = \Lambda_T \delta_{\beta}X^{\mu}$. In order to ensure that the pullback field $B_i$ varies under $\delta_T$ as above, we can compensate for the absence of dynamical $X$'s by endowing the target space $B_{\mu}$ with the transformation
\begin{align}	
\begin{split}
	\delta_{T}(\delta_i^\mu B_{\mu})&=\partial_{i}({\Lambda}_{T}\beta^k)\delta_k^\mu B_{\mu}+\Lambda_{T}\beta^{k}\partial_k(\delta_i^{\mu} B_{\mu}) \,.
\end{split}
\end{align}
If we now make the replacements $B_i \to B_i^{(A)}$ as in \eqref{E:BtoBA}  and modify the measure:  $\sqrt{-g} \to \sqrt{-g}/(\beta^iA_i+1)$, then a short computation shows that $\delta_TS= 0$ implies
\begin{equation}
\label{E:entropyprobe}
	\partial_i S^i  = \beta^iG_{ij}J^j-\nu \partial_i J^i.
\end{equation}

Note that the Ward identity for $\delta_T$ in the probe limit~\eqref{E:entropyprobe} is not quite the Ward identity in the full sigma model~\eqref{E:bosonicsigma}. The identity for the probe limit is missing the term $-\beta_i \partial_j T^{ij}$, essentially because we lack dynamical $X$'s. Put differently, in the probe limit we are computing only the contribution of the $B_i$'s to $S^i$. The contribution of the thermal background and the backreaction of the background to $S^i$ is not present. Thus, there is no term in $S^i$ which will compensate for the generation of the Joule heating term on the right-hand side of \eqref{E:entropyprobe}. We note that $S^i$ will be conserved on-shell if we set the external field $B_{\mu}$ such that $\beta^iG_{ij}=0$.

We are now ready to construct the super entropy current. As was the case for the bosonic sigma model described above, we would like to gauge the $\delta_{\beta}$ transformation described in \eqref{E:commrelation} so that a conserved super entropy current will emerge.  Since we are currently working in the probe limit where the $X$'s have been replaced by their classical on-shell values, we can no longer require that, say, $\delta_T X_1^{\mu} = \Lambda_T \delta_{\beta} X_1^{\mu}$. We may only impose
\begin{equation}
\label{E:superCtrans}
	\delta_{T} \super{C} = \super{\Lambda}_T \delta_{\beta}\super{C}\,,
\end{equation}
where now the ``gauge'' parameter may be a superfield. As in our discussion of the probe limit of the bosonic sigma model we endow the target space gauge field with the following transformation law
\begin{equation}	
	\delta_{T}(\delta_i^\mu\SB_{\mu})=\partial_{i}(\super{\Lambda}_{T}\beta^k)\delta_k^\mu\SB_{\mu}+\super{\Lambda}_{T}\beta^{k}\partial_k(\delta_i^{\mu}\SB_{\mu})
\end{equation}
where
\begin{equation}
\SB_{\mu}=\mathcal{R}\left(\frac{B_{1\mu}(X_{eq})+B_{2\mu}(X_{eq})}{2}\right)+\btheta\theta\mathcal{A}\left(B_{1\mu}(X_{eq})-B_{2\mu}(X_{eq})\right)\,,
\end{equation}
and $X_{eq}^{\mu}$ is given in \eqref{E:probeX}.
With this transformation, we ensure that
\begin{equation}
	\delta_T \super{B}_i^{(A)} = \super{\Lambda}_T\delta_{\beta}\super{B}_i^{(A)}\,,
	\qquad
	\delta_T \eta_{ij} = 0\,,
\end{equation}
where
\begin{equation}
\label{E:superreplacement}
	\SB_{i}^{(A)}=\SB_i+\SA_i\bbnu 
\end{equation}
and we have defined
\begin{equation}
	\bbnu = \beta^i \super{B}_i + \Lambda_{\beta}\,.
\end{equation}

The replacements
\begin{equation}
\label{E:superreplacement}
	\SB_i \rightarrow \SB_{i}^{(A)}\,, \qquad 
	\super{D}_i \to \super{D}_{(\super{A})\,i} + \super{A}_i \delta_{\beta} 
\end{equation}
now almost ensure that the Lagrangian $L$ defined in \eqref{E:actionexpansionshort} satisfies $\delta_T L = \super{\Lambda}_T \delta_{\beta}L$. The problematic terms are those that contain superderivatives, e.g., ${D}_{\theta}\super{B}_i$. In order for $L$ to transform correctly under $\delta_{T}$, one needs the additional substitutions
\begin{equation}
\label{E:Dthetadef}
	{D}_{\theta} \to    \frac{\partial}{\partial \theta} -i \delta_{\beta}\btheta + \super{A}_{\theta} \delta_{\beta}\,,
	\qquad
	{D}_{\btheta} \to  \frac{\partial}{\partial \btheta} + \super{A}_{\btheta} \delta_{\beta} \,, \\
\end{equation}
together with the transformation properties
\be\label{eq:transA}
	\delta_{T} \super{A}_{\theta} = \super{\Lambda}_T \delta_{\beta} \super{A}_{\theta} - \super{A}_{\theta} \delta_{\beta} \super{\Lambda}_T - {D}_{\theta} \super{\Lambda}_T
\ee
and an analogous expression for $\delta_T \super{A}_{\btheta}$. The substitution \eqref{E:Dthetadef} implies that we extend the connection $\super{A}_{i}$ to a superconnection $\super{A}_I$. The authors of \cite{Haehl:2015foa,Haehl:2015uoc} have entertained the possibility that $\super{A}$ is a dynamical gauge field which condenses and that the condensate is associated with the breaking of a worldvolume CPT. 

Let us study what changes when we have a super-connection $\super{A}_I$ rather than an ordinary connection $A_i$. To illustrate this as simply as possible, we start with the original action $\int d\theta d\btheta \, L$ and neglect its KMS partner. We take this action and rescale the measure so that the integral is invariant under $\delta_T$. This action becomes $S_0 = \int d\theta d\btheta\, \super{M} L$ with
\beq
	\super{M}=\frac{\sqrt{-\super{g}}}{\beta^i\super{A}_i+1}\,.
\eeq
Define the flavor ``supercurrent'' $\super{j}^i$ conjugate to $\super{B}_i$ and the ``supercurrent'' ${\ssigma}^I$ conjugate to $\super{A}_I$ as
\beq
	\delta S_0 = \int d\theta d\btheta \, \super{M}\left( \super{j}^i \delta \super{B}_i - \ssigma^I \delta \super{A}_I\right)\,.
\eeq
Then $\delta_T S_0=0$ implies that $\ssigma^I$ satisfies, when $\super{A}_I = 0$,
\begin{equation}\label{E:identity}
	\super{D}_i \ssigma^i + {D}_{\theta} \ssigma^{\theta} + {D}_{\btheta} \ssigma^{\btheta} = - \bbnu(\mathcal{A}^{-1}{\cal R}^{-1}\partial_i \super{j}^i)+(\beta^{k}\super{G}_{ki})(\mathcal{A}^{-1}{\cal R}^{-1}\super{j}^i)\,.
\end{equation}
Expanding $\ssigma^I$ in components,
\begin{equation}
	\ssigma^I = \sigma_{b}^{ I} + \theta \sigma_{\bar{g}}^I + \btheta \sigma_{g}^I +\btheta\theta \sigma_{t}^I
\end{equation}
then the bottom component of  this  identity  reads
\begin{equation}
\label{E:sigmanon}
	\partial_i \sigma_b^i\big|_{\super{A}_I=0} = -\sigma_{\bar{g}}^{\theta} - \sigma_g^{\btheta} -(\mathcal{R}\nu_r)(\mathcal{A}^{-1}\partial_i j_r^i) + (\mathcal{R}\beta^{k}G_{r\,ki})(\mathcal{A}^{-1}j_r^i)\,,
\end{equation}
with $ \super{j}^i \big|_{\theta=\btheta=0}={\cal R}j_r^i $ and $\super{G}_{ij}\big|_{\theta=\btheta=0}={\cal R}G_{r\,ij}$.

On shell, the right-hand side of \eqref{E:sigmanon} is nonzero due to the Joule heating term as well as the $\sigma_{\bar{g}}^{\theta}$ and $\sigma_g^{\btheta}$ terms. The latter two may be computed either from varying the action with respect to $\super{A}_{\theta}$ and $\super{A}_{\btheta}$, or by the following. If we were to drop $\super{A}_{\theta}$ and $\super{A}_{\btheta}$ by hand from the tensor terms, thereby making the tensor terms non-invariant under $\delta_T$, then $\sigma_{\bar{g}}^{\theta}+\sigma_g^{\btheta}$ is the variation of the tensor term under $\delta_T$. Either way, one obtains
\begin{equation}
	-\sigma_{\bar{g}}^{\theta} - \sigma_g^{\btheta} = -\frac{1}{2} L^{ij}(D_k)\delta_{\beta}({\cal R}B_{r\,i})\,\delta_{\beta}({\cal R}B_{r\,j})\,.
\end{equation}

Let us now turn our attention to the full action \eqref{E:fullaction}, including the KMS partner term. Adding a connection $\super{A}_i$ to ordinary derivatives and substituting
\begin{equation}
	\sqrt{-\super{g}} \to \super{M} \equiv \frac{\sqrt{-\super{g}}}{\beta^i \super{A}_i+1}\,, \\
	\qquad
	\sqrt{-\wt{\super{g}}} \to \wt{\super{M}} \equiv \frac{\sqrt{-\wt{\super{g}}}}{\beta^i {\super{A}}_i+1}\,,
\end{equation}
we may define
\begin{equation}
\label{E:superN}
	\super{S}^i \equiv -\frac{1}{\super{M}} \frac{\partial{S_{eff}}}{\partial \super{A}_i}\,.
\end{equation}
Setting ghost terms and $\SA$ to zero, the variation of the action under $\super{\Lambda}_T = \btheta \theta \mathcal{A}\Lambda_T$ yields
\begin{equation}\label{eq:varSeff}
\delta_{T}S_{eff}=\int d^{d}\sigma\mathcal{A}\Lambda_{T}\left(-\partial_{i}S^{\prime i}-S_{\bar{g}}^{\theta}-S_g^{\btheta}-(\mathcal{R}\nu_r)(\mathcal{A}^{-1}\partial_{i}J_r^{i})+(\mathcal{R}\beta^kG_{r\,ki})(\mathcal{A}^{-1}J_r^{i})\right)\,,
\end{equation}
 where $S^{\prime i} =\super{S}^i\big|_{\theta=0,\,\btheta=0}$ and $S_g^{\btheta} + S_{\bar{g}}^\theta$ specifies the non-invariance of the tensor terms under $\delta_T$ if we were to drop $\super{A}_{\theta}$ and $\super{A}_{\btheta}$. Thus, the off-shell divergence of $S^{\prime i}$ is given by
\begin{equation}\label{eq:U1ward}
	\partial_{i} S^{\prime i} \Big|_{\super{A}_{I}=0} =
	-{S}^{\theta}_{\bar{g}} -{S}^{\btheta}_g- (\mathcal{R}\nu_r)(\mathcal{A}^{-1}\partial_i J_r^i)+(\mathcal{R}\beta^{k}G_{r\,ki})(\mathcal{A}^{-1}J_r^i)\,,
\end{equation}
The rightmost term in equation \eqref{eq:U1ward} is associated with the Joule heating term. Observe the similarity with the off-shell entropic second law~\eqref{E:2ndS} up to the presence of $\mathcal{R}$ and $\mathcal{A}$.

Note that  since we have obtained $\super{S}^I$ from a variational principle, we can now define it quantum mechanically by taking an appropriate variation of the Schwinger-Keldysh generating function. Thus, we can treat \eqref{eq:U1ward} as a Ward identity for the divergence of $S'^i$.

In the remainder of this work we will consider the tree level expression for $S'^i$ and relate it to the hydrodynamic entropy current $S^i$. More precisely, we will relate it to the contribution of the charge to the entropy current. We first show that $S'^i - T_{eq}^{ij}\beta_j =su^i+\mathcal{O}(\partial)$, where $s$ is the entropy density. Here $T_{eq}^{ij}$ is the pullback of the equilibrated stress tensor. Its appearance is a result of the probe limit we are working in---the effective action only captures the dynamics of  the charge and the $T_{eq}^{ij} \beta_j$ term provides for the contribution of the thermal background to the entropy density.

We will then argue that when we set the electric field to zero, $\beta^i G_{ij}=0$, the right-hand side of \eqref{eq:U1ward} is non-negative up to total derivatives, when working perturbatively in the derivative expansion and placing all fields on-shell. (The total derivative terms come in at fourth order in the derivative expansion.) As we will show explicitly this implies that we may add terms to the consistent entropy current ${S'}^i$ to generate a ``hydrodynamic'' entropy current $S^i$ which must satisfy \eqref{E:localsecond} and \eqref{E:localconstitutive}. We thereby identify the total entropy production to be the spacetime integral of $-S_{\bar{g}}^{\theta} - S_g^{\btheta}$.

If a saddle point approximation exists we can evaluate the entropy current by varying the effective action with respect to the sources to obtain the tree level expression for $\super{S}^{\prime I}$. Let us start with the ungauged effective action expanded at zeroth order in derivatives. We find 
\begin{equation}
\label{E:Sideal}
	S_{eff} = \int d^d\sigma d\theta d\btheta\, \super{M}\left(\frac{1}{2}{F}(T,\,\bbnu)+\frac{1}{2}{F}(T,-\bbnu)\right) + \mathcal{O}(\partial)\,,
\end{equation}
up to boundary terms,
where
\begin{equation}
\label{E:superthermal}
	 T=\left(-\beta^i\beta^j\eta_{ij}\right)^{-1/2} \,,\quad\hbox{and}\quad 
	 \bbnu= \left(\beta^i\super{B}_i+\Lambda_{\beta}\right)  \,.
\end{equation}
Defining
\begin{equation}
	P(T,\bbnu) \equiv \frac{1}{2}{F}(T,\,\bbnu)+\frac{1}{2}{F}(T,-\bbnu) 
\end{equation}
the equilibrium constitutive relations take the form
\begin{equation}
	\super{J}^i = \left(\frac{\partial P}{\partial \bbnu}\right)_{T} {\beta}^i + \mathcal{O}(\partial)\,.
\end{equation}
Since the bottom component of $\super{J}^i$ should be identified with the charge current, we identify $P$ with the pressure, $T$ with the temperature, $\beta^i$ with a normalized velocity $u^i = T \beta^i$ and the bottom component of $\bbnu$ with $\mu/T$ where $\mu$ is the chemical potential. 

The version of~\eqref{E:Sideal} invariant under $\delta_T$ is
\begin{equation}
S_{eff}=\int  \frac{d^d\sigma d\theta d\btheta}{\beta^{i}\SA_i+1}P\left(T,\bbnu^{(A)}\right)\,,
\end{equation}
where 
\begin{equation}
 \bbnu^{(A)}=(1+\beta^i\SA_i)\bbnu\,.
\end{equation}
Varying the action with respect to $\super{A}_i$ and taking the bottom component, we find that
\begin{equation}
	S'^i -T_{eq}^{ij}\beta_{j} =\Big(\frac{\epsilon+P}{T}-\frac{\nu}{T}\left(\frac{\partial P}{\partial \nu}\right)_T\Big)u^i+\mathcal{O}(\partial)=su^i+\mathcal{O}(\partial)
\end{equation}
where  the last equality follows from the constitutive relation for the stress tensor in equilibrium, $T^{ij}_{eq} = \epsilon u^i u^j +P (\eta^{ij} + u^i u^j)$, the first law, $dP=sdT+\rho d\mu$, and the Gibbs-Duhem relation, $\epsilon+P=sT+\rho\mu$.

Next consider the right-hand side of \eqref{eq:U1ward}. After a straightforward but somewhat tedious computation, we find 
\begin{equation}
\label{E:divergence}
	-S^{\theta}_{\bar{g}} - S^{\btheta}_g =
	-\frac{1}{2} L^{ij}(D_k)\delta_{\beta}({\cal R}B_{r\,i})\,\delta_{\beta}({\cal R}B_{r\,j}) -\frac{1}{2}\eta\,\delta_{\beta} \mathcal{R} B_{r\,i}  {\cal S}^{-1}\left( {L}^{ij}(\eta_{\partial}D_k) 
	 \delta_{\beta}({\cal R}{\cal S}  B_{r\,i})  \right)\,.
\end{equation}
where we have used $\mathcal{S}=\frac{1}{2}(1+e^{-i\delta_{\beta}})$ and $\eta$ is the CPT eigenvalue of $L^{ij}$. 
We have also set all but the bottom components of $\super{B}$ to zero and have omitted the dependence of $L^{ij}$  on $\beta^i$, $\eta_{ij}$ and $\Lambda_{\beta}$.

Recall that unitarity implies
\begin{equation}
\label{E:PositiveSeff}
	\hbox{Im}( S_{eff} ) \geq 0
\end{equation}
(see \cite{Glorioso:2016gsa}).
As emphasized by \cite{Glorioso:2016gsa,Glorioso:2017fpd} it is difficult to constrain a Lagrangian so that \eqref{E:PositiveSeff} is satisfied. In particular, one can add total derivatives to the effective Lagrangian while keeping $S_{eff}$ unchanged. However, the off-shell constraint \eqref{E:PositiveSeff} must be satisfied for any field configuration. In particular, it should be satisfied for configurations where the dynamical fields and sources are constant. It then follows that the imaginary part of the effective Lagrangian must be a positive function when neglecting derivatives. See Appendix F of \cite{Glorioso:2016gsa} for a detailed discussion. 
Evaluating the imaginary part of the action in the absence of derivatives, we find that
\begin{equation}
\label{E:Imleff2}
	\hbox{Im}\, L_{eff}\Big|_{\partial=0} = \sigma^{ij}  B_{a\,i} B_{a\,j} +{\cal O}(B_a^4)\,.
\end{equation}
where
$
	\sigma^{ij} = -\left( L^{ij} + \eta\wt{L}^{ij}\right)\Big|_{\partial=0}
$ and $\wt{L}$ refers to the KMS-conjugate Lagrangian.
The expression in \eqref{E:Imleff2} must be positive for all values of $B_{a}$ in general, and for small $B_{a}$ in particular. Thus, $\sigma^{ij}$ must be non negative. In what follows we will assume that it is strictly positive and therefore invertible.\footnote{In practice, we may use a change of fluid frame to modify certain components of $\sigma^{ij}$ to vanish. Such a change of frame may be carried out order by order in the derivative expansion and will not affect the argument below.
}

Given that $\left| \left| \sigma^{ij}\right| \right| > 0$, it follows that the right-hand side of \eqref{E:divergence} must also be positive at leading order in derivatives. Furthermore, at subleading order in derivatives the right-hand side of \eqref{E:divergence} will always include at least two factors of $\delta_{\beta}B_r$ or their derivatives. Thus, a term with $n+1$ derivatives on the right-hand side of \eqref{E:divergence} may always be brought into the form $\delta_{\beta} B_{r\,i} Q^{i}_{(n)}$ up to total derivatives, with $Q^{(i)}_{(n)}$ a term with $n$ derivatives \cite{Bhattacharyya:2013lha,Bhattacharyya:2014bha,Haehl:2015pja,Glorioso:2016gsa,Glorioso:2017fpd}. Thus, we may always write
\begin{equation}
	-S^{\theta}_{\bar{g}} -S^{\btheta}_g = 
	\sigma \left(\delta_{\beta}B_{r} + \frac{1}{2}\sigma^{-1} \left(Q_{(2)} + \ldots Q_{(n-1)}\right) \right)^2
	+ \partial J_S
\end{equation}
where we have omitted flavor and spacetime indices for brevity. Thus, if we define
\begin{equation}
	S^i = {S'}^i - J_S^i\,,
\end{equation}
then $S^i$ satisfies both \eqref{E:localsecond} and \eqref{E:localconstitutive} and is therefore the (hydrodynamic) entropy current.

%------------------------------------------------------
\section{Entropy production}
\label{S:newentropy}
%------------------------------------------------------

While we have shown that the hydrodynamic entropy current can be constructed so that it has non-negative divergence at any order in the derivative expansion, one may inquire about positivity of entropy production in general. In this Section we show that the total entropy increases, to quadratic order in fields but independent of a derivative expansion. This analysis complements that of \cite{Glorioso:2016gsa} where a similar statement was made in the statistical mechanical limit.

Since the entropy current analysis is carried out to quadratic order in the fields it is convenient for our current purpose to write the full effective action after superspace integration, in momentum space. Such a construction was carried out in \cite{Crossley:2015evo}. We rederive it here for completeness. Let us consider the pulled-back fields $F_{r\,i} = \frac{1}{2} \left(B_{1\,i}+B_{2\,i}\right)$ and $F_{a\,i} = B_{1\,i}-B_{2\,i}$. We define the Fourier transform of these fields as
\beq
	\obar{F}(\omega,\vec{k}) =\int d\sigma^0d^{d-1}\vec{\sigma} \, e^{-i \omega t + i \vec{k}\cdot \vec{x}} F(\sigma^0,\vec{\sigma})\,.
\eeq
The most general local effective action quadratic in the fields will contain terms proportional to $F_r^2$, $F_r F_a$ and $F_a^2$. The Schwinger-Keldysh 
%topological 
symmetry ensures that correlation functions of all $a$-type fields must vanish. Thus, the effective action must not contain any terms quadratic in the $F_r$'s, 
\beq
	\label{E:Seff}
	S_{eff} = \int \frac{d\omega d^{d-1}k}{(2\pi)^d} \left\{ \obar{G}_R^{ij}(\omega,\vec{k}) \obar{F}_{r\,i}(\omega,\vec{k}) \obar{F}_{a\,j}(-\omega,-\vec{k}) + \frac{1}{2}\obar{G}_S^{ij}(\omega,\vec{k}) \obar{F}_{a\,i}(\omega,\vec{k}) \obar{F}_{a\,j}(-\omega,-\vec{k})\right\}\,.
\eeq

If the $F_i$ are external fields, then $-i\obar{G}_R^{ij}$ and $\obar{G}_{S}^{ij}$ are, respectively, the Fourier transformed retarded and symmetrized two point functions for the current. While we have focused on the probe limit in this text we could, as argued in \cite{Glorioso:2016gsa}, discuss more general fields if we restrict ourselves to the quadratic part of the action. In this case the $F_{r\,i}$ and $F_{a\,j}$ would correspond to $r$ and $a$-type fields associated with the sources on which the generating functional depends and $i$ and $j$ would be indices appropriate to that source. In what follows we will keep the $i$ and $j$ indices but the reader should keep in mind that these may not necessarily refer to gauge fields.

The reality condition on $S_{eff}$ is given by \cite{Crossley:2015evo}
\beq
	\big( S_{eff}(F_{r\,i},F_{a\,j})\big)^* = - S_{eff}(F_{r\,i},-F_{a\,j})\,,
\eeq
which implies that $\obar{G}_R$ and $\obar{G}_S$ satisfy
\begin{align}
\begin{split}
	\label{E:Greal}
	\left(\obar{G}_R^{ij}(\omega,\vec{k})\right)^* &= \obar{G}_R^{ij}(-\omega,-\vec{k})\,,
	\\
	\left( \obar{G}_S^{ij}(\omega,\vec{k})\right)^* & = -\obar{G}_S^{ij}(-\omega,-\vec{k})\,,
\end{split}
\end{align}
Because $\obar{G}_S^{ij} = \obar{G}_S^{ji}$ and $\obar{F}_{a\,i}(-\omega,-\vec{k}) = \left( \obar{F}_{a\,i}(\omega,\vec{k})\right)^*$, then this together with the positivity condition $\text{Im}(S_{eff})\geq 0$ implies that $-i\obar{G}_S^{ij}$ is a hermitian, positive semi-definite matrix.

In thermal states there is a second topological symmetry, the KMS topological symmetry, which is the statement that correlation functions of the $\tilde{a}$-type operators vanish. Working in the static gauge (see Footnote \ref{F:staticgauge}) it implies 
\beq
	\label{E:KMSReG}
	\text{Re}\left( \obar{G}_R^{ij}(\omega,\vec{k})\right) = \text{Re}\left( \obar{G}_R^{ji}(\omega,\vec{k})\right)\,,
\eeq
and
\beq
	\label{E:FDT}
	\obar{G}_S^{ij}(\omega,\vec{k}) = \frac{i \coth\left( \frac{\beta \omega}{2}\right)}{2}\text{Im}\left(\obar{G}_R^{ij}(\omega,\vec{k}) + \obar{G}_R^{ji}(\omega,\vec{k})\right)\,.
\eeq
Note that~\eqref{E:FDT} is consistent with the fact that $-i \obar{G}_S^{ij}$ is hermitian. 

Note that \eqref{E:FDT} bears a striking resemblance to the fluctuation dissipation theorem, whereby the symmetrized Green's function is determined by the retarded one. This resemblance is not accidental. The entirety of our analysis so far may be directly transferred over the the Schwinger-Keldysh generating functional of connected correlators $W$. Further, in that case the imaginary part of $\obar{G}_R^{ij}+\obar{G}_R^{ji}$ is the matrix of spectral functions, which unitarity implies is also positive semi-definite at positive frequency
\beq
	\left|\left| \text{Im}\left(\obar{G}_R^{ij}+\obar{G}_R^{ji}\right)\right|\right| \geq 0\,, \qquad \omega \geq 0\,,
\eeq
and negative semi-definite at negative frequency. It follows from~\eqref{E:FDT} that $-i\obar{G}_S^{ij}$ is positive semi-definite for all $\omega$ as it should be. Finally, having accounted for these properties, the full KMS symmetry imposes certain discrete transformation laws of the various components of $\obar{G}_R^{ij}$ under CPT. 

Putting all of the pieces together, a more useful way to characterize the effective action is in terms of the real and imaginary parts of $\obar{G}_R$,
\begin{multline}
	\label{E:Seffv2}
	S_{eff} = \int \frac{d\omega d^{d-1}k}{(2\pi)^d}\left\{ \frac{\text{Re}\left(\obar{G}_R^{ij}(\omega,\vec{k})\right)}{2} 
	\left( \obar{F}_{1\,i}(\omega,\vec{k})\obar{F}_{1\,j}(-\omega,-\vec{k}) - (1\leftrightarrow 2)\right) \phantom{\frac{\left(\obar{G}_R^{ij}\right)}{\beta\omega} }\right.
	\\
	\qquad \qquad \qquad \qquad\qquad  \left.+ \frac{i\,\text{Im}\left(\obar{G}_R^{ij}(\omega,\vec{k})\right)}{1-e^{-\beta\omega}} {\obar{\wt{F}}}_{a\,i}(\omega,\vec{k}) \obar{F}_{a\,j}(-\omega,-\vec{k})\right\}\,,
\end{multline}
where $\obar{\wt{F}}_{a\,i}$ is the Fourier-transform of $\wt{F}_{a\,i}$, which in terms of the average and difference combinations is
\beq
	\label{E:Ftilde}
	\obar{\wt{F}}_{a\,i}(\omega,\vec{k}) = (1-e^{-\beta\omega})\left(  \obar{F}_{r\,i}(\omega,\vec{k}) + \frac{1}{2}\coth\left( \frac{\beta\omega}{2}\right)\obar{F}_{a\,i}(\omega,\vec{k})\right)\,.
\eeq

Let us demonstrate that~\eqref{E:Seffv2} is equivalent to the original expression~\eqref{E:Seff} for $S_{eff}$. First, using~\eqref{E:KMSReG} we find that the first term in the effective action equals
\begin{multline}
\label{E:12tora}
	\int \frac{d\omega d^{d-1}k}{(2\pi)^d} \frac{\text{Re}\left( \obar{G}_R^{ij}(\omega,\vec{k})\right)}{2} \left( \obar{F}_{1\,i}(\omega,\vec{k}) \obar{F}_{1\,j}(-\omega,-\vec{k})- (1\leftrightarrow 2)\right) 
	\\
	=\int \frac{d\omega d^{d-1}k}{(2\pi)^d} \text{Re}\left( \obar{G}_R^{ij}(\omega,\vec{k})\right)\obar{F}_{r\,i}(\omega,\vec{k})\obar{F}_{a\,j}(-\omega,-\vec{k})\,.
\end{multline} 
One can recover \eqref{E:Seff} from \eqref{E:Seffv2} by inserting \eqref{E:12tora} as well as the expression~\eqref{E:Ftilde} for $\obar{\wt{F}}_{a}$ into the effective action \eqref{E:Seffv2} with $\obar{G}_S$ determined as in \eqref{E:FDT}.

The virtue of~\eqref{E:Seffv2} is that it more readily manifests the symmetries of the problem. The first term, being of the form $S_1 - S_2$, manifestly respects the Schwinger-Keldysh and full KMS symmetries.The second, being linear in the $a$-type fluctuations, automatically respects the Schwinger-Keldysh symmetry. It is also invariant under the combination of CPT and exchanging the $a$- and $\tilde{a}$-fluctuations, and so respects the full KMS symmetry.

It is a little tricky to couple the fields in the effective action to the external field $A$ and so deduce the conjugate current. However it is straightforward to deduce the entropy production, which in the previous section was given by the integral of $-S_{\bar{g}}^{\theta} - S_g^{\btheta}$. Without introducing ghosts, the transformation we studied in the previous Section, $\delta_T$ with $\super{\Lambda}_T = \btheta \theta \mathcal{A} \Lambda_T$, is no longer a symmetry of the effective action, and the entropy production is simply the non-invariance under it
\beq
\label{E:deltaT}
	\delta_T F_{r\,i} = 0\,, \qquad \delta_T (\mathcal{A} F_{a\,i}) = \Lambda_T\mathcal{R} \beta \partial_t F_{r\,i}\,,
\eeq
with $\Lambda_T$ a constant. The total entropy production $\Delta S$ is given by $\Delta S =-\left. \frac{\partial (\delta_T S_{eff})}{\partial \Lambda_T}\right|_{F_a=0}$. The variation of the first term in the effective action~\eqref{E:Seffv2} proportional to $\text{Re}\left(\obar{G}_R^{ij}\right)$ is
\beq
	\delta_T  \text{Re}(S_{eff}) = 2i\Lambda_T  \int \frac{d\omega d^{d-1}k}{(2\pi)^d} \text{Re}\left( \obar{G}_R^{ij}(\omega,\vec{k})\right)\tanh\left( \frac{\beta\omega}{2}\right)  \obar{F}_{r\,i}(\omega,\vec{k}) \obar{F}_{r\,j}(-\omega,-\vec{k})\,.
\eeq
where we have used that $\mathcal{R}/\mathcal{A} = \frac{2}{i\beta \partial_t}\tanh\left( \frac{i\beta\partial_t}{2}\right)$. This variation vanishes: since $\text{Re}\left(\obar{G}_R^{ij}(\omega,\vec{k})\right)$ is symmetric under $i\leftrightarrow j$ as well as under $(\omega,\vec{k})\to (-\omega,-\vec{k})$, while the rest of the the integrand is odd under the combination of those two transformations. 

The variation of the second term proportional to $\text{Im}\left( \obar{G}_R^{ij}\right)$ is 
\beq
	\delta_T \text{Im}(S_{eff}) = -2\Lambda_T  \int \frac{d\omega d^{d-1}k}{(2\pi)^d} \text{Im}\left(\obar{G}_R^{ij}(\omega,\vec{k})\right) \tanh\left( \frac{\beta\omega}{2}\right) \obar{F}_{r\,i}(\omega,\vec{k})\obar{F}_{r\,j}(-\omega,-\vec{k})\,.
\eeq
Only the symmetric part of $\text{Im}\left( G_R^{ij}\right)$ contributes to this variation, and using $\obar{F}_{r\,j}(-\omega,-\vec{k}) = \obar{F}_{r\,j}^*(\omega,\vec{k})$ we find
\beq
	\label{E:DeltaS1}
	\Delta S =\int \frac{d\omega d^{d-1}k}{(2\pi)^d}\tanh\left( \frac{\beta\omega}{2}\right) \text{Im}\left( \obar{G}_R^{ij}(\omega,\vec{k})+\obar{G}_R^{ji}(\omega,\vec{k})\right) \obar{F}_{r\,i}(\omega,\vec{k}) \obar{F}_{r\,j}^*(\omega,\vec{k})\,.
\eeq
Using ~\eqref{E:FDT} we rewrite this as
\beq
	\label{E:DeltaS2}
	\Delta S= 2\int \frac{d\omega d^{d-1}k}{(2\pi)^d}  \tanh^2\left( \frac{\beta\omega}{2}\right) \left(-i \obar{G}_S^{ij}(\omega,\vec{k})\right) \obar{F}_{r\,i}(\omega,\vec{k})\obar{F}_{r\,j}^*(\omega,\vec{k})\,.
\eeq
The integrand is positive on account of $-i \obar{G}_S^{ij}$ being a symmetric, real, positive semi-definite matrix, so we find
\beq
	\Delta S\geq 0\,,
\eeq
as expected.

We wrap up with two brief comments. First, we may take variations of the on-shell entropy with respect to the external fields and obtain correlation functions of the entropy production with the currents, giving
	\beq
	\langle \Delta S \, \obar{J}^i_a(\omega,\vec{k}) \obar{J}^j_a(-\omega,-\vec{k})\rangle =2 \tanh\left( \frac{\beta\omega}{2}\right)\text{Im}\left( \obar{G}_R^{ij}(\omega,\vec{k}) + \obar{G}_R^{ji}(\omega,\vec{k})\right)\,.
	\eeq
Second, observe that only the symmetric part of $\text{Im}(\obar{G}_R^{ij})$ contributes to the entropy production. The real part of $\obar{G}_R^{ij}$ and the antisymmetric part of $\text{Im}(G_R^{ij})$ do not. This generalizes a known result in hydrodynamics. For the effective action describing relativistic hydrodynamics, the pressure term contributes to $\text{Re}(\obar{G}_R)$, while the leading contribution to the symmetric part of $\text{Im}(\obar{G}_R)$ is the ordinary conductivity. Relatedly, in two spatial dimensions the leading contribution to the antisymmetric part of $\text{Im}(\obar{G}_R)$ is the anomalous Hall conductivity, which is also known to be dissipationless~\cite{Jensen:2011xb}.

%------------------------------------------------------
\begin{acknowledgments}
%------------------------------------------------------

We would like to thank J.~Cotler, P.~Glorioso, H. Liu, and V.~Mikhaylov for discussions. The work of KJ was supported in part by the US Department of Energy under Grant No. DE-SC0013682. The work of NPF in KU Leuven was supported in part by the National Science Foundation of Belgium (FWO) grant G.001.12 Odysseus and by the European Research Council grant no. ERC-2013-CoG 616732 HoloQosmos. The work of NPF was also supported in part by the Israel Science Foundation under grant 504/13 and in part at the Technion by a fellowship from the Lady Davis Foundation. The work of AY and RM was supported in part by the Israeli Science Foundation under an ISF-UGC grant 630/14 and an ISF excellence center grant 1989/14.

%------------------------------------------------------
\end{acknowledgments}
%------------------------------------------------------

\bibliography{entropy_biblio}{}
\bibliographystyle{JHEP}

\end{document}